\documentclass{PoS}

\def\321{$\mathrm{SU(3) \otimes SU(2) \otimes U(1)}$ }
\def\lfv{lepton flavor violation }
\def\clfv{charged lepton flavor violation }

\def\lnv{lepton number violating }

\def \znbb {$\rm 0\nu\beta\beta$ }

\def\gsim{\raise0.3ex\hbox{$\;>$\kern-0.75em\raise-1.1ex\hbox{$\sim\;$}}}
\def\lsim{\raise0.3ex\hbox{$\;<$\kern-0.75em\raise-1.1ex\hbox{$\sim\;$}}}
\def\lfv{lepton flavor violation }
\def\lnv{lepton number violating }
\def\SM{$\mathrm{SU(3)_c \otimes SU(2)_L \otimes U(1)_Y}$ }
\newcommand{\sm}{{Standard Model }}

\def\gsim{\raise0.3ex\hbox{$\;>$\kern-0.75em\raise-1.1ex\hbox{$\sim\;$}}}
\def\lsim{\raise0.3ex\hbox{$\;<$\kern-0.75em\raise-1.1ex\hbox{$\sim\;$}}}

\definecolor{linkcolor}{rgb}{0,0,0.5}

 \usepackage[usenames,dvipsnames]{xcolor}
 \usepackage[normalem]{ulem}

\title{Neutrino physics outlook}

\ShortTitle{Neutrino physics outlook}

\author{{Jos\'e W. F. Valle}
\thanks{ This work was supported by the Spanish grants SEV-2014-0398 and FPA2017-85216-P (AEI/FEDER, UE), PROMETEO/2018/165 (Generalitat Valenciana) and the Spanish Red Consolider MultiDark FPA2017-90566-REDC. }\\
\textcolor{black}{AHEP Group, Institut de F\'{i}sica Corpuscular --   C.S.I.C.\\
Universitat de Val\`{e}ncia, Parc Cientific de Paterna.\\   C/Catedratico Jos\'e Beltr\'an, 2 E-46980 Paterna (Val\`{e}ncia) - SPAIN}\\
        E-mail: \email{valle@ific.uv.es}}

\abstract{Here I identify some of the opportunities in particle physics associated to the lepton sector, according to their perceived significance, and taking into account the current state-of-the-art.}

\FullConference{The 21st international workshop on neutrinos from accelerators (NuFact2019)\\
August 26 - August 31, 2019\\
Daegu, Korea}

\begin{document}
\section{Neutrino oscillations}

\begin{figure}[h]
    \centering
\includegraphics[scale=.4]{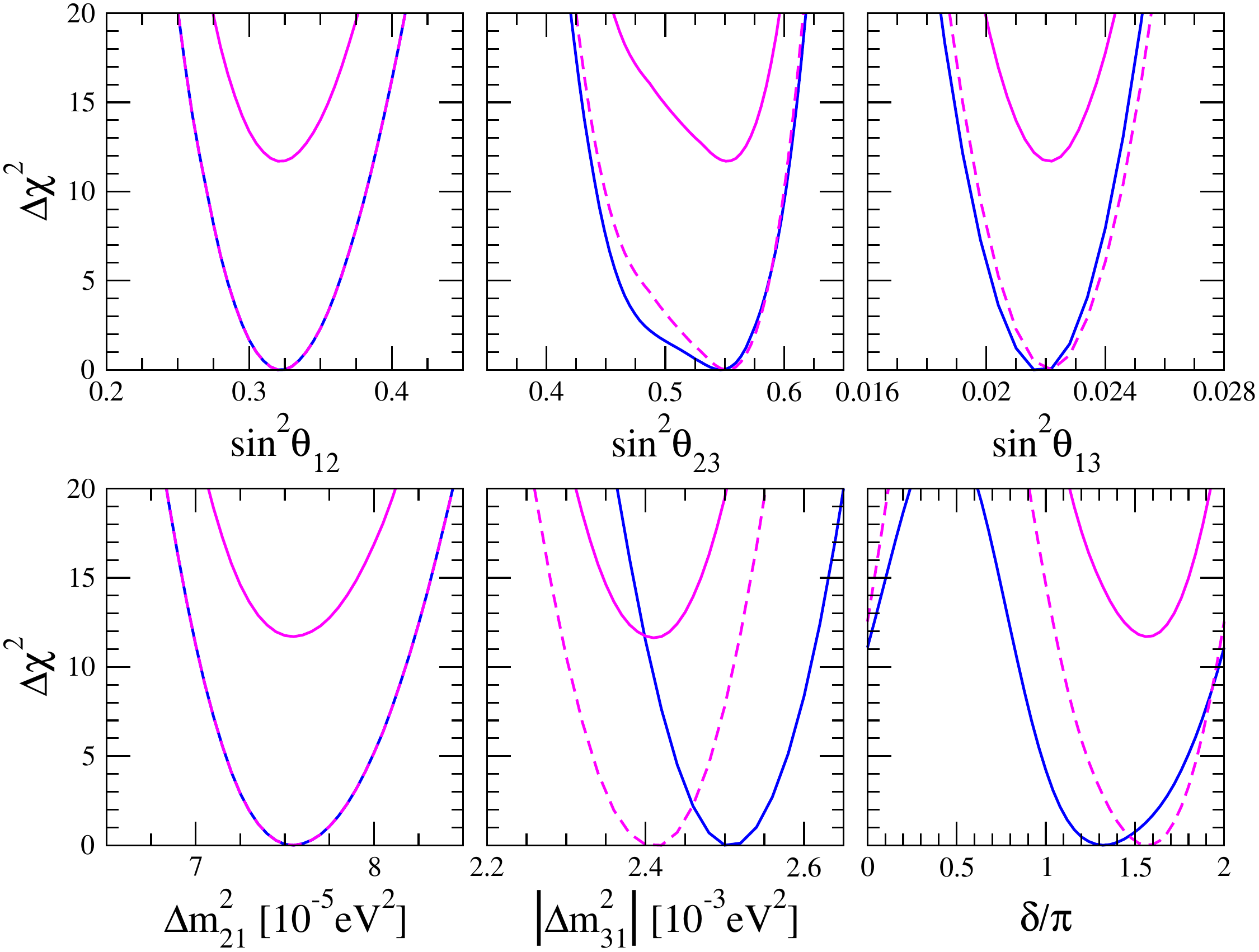}
 \caption{Summary of neutrino oscillation parameters, 2018. From~\cite{deSalas:2017kay}.}
    \label{fig:osc}
  \end{figure}

My assigned task was to give an outlook on the field.
I will take the opportunity to discuss some of the most interesting directions for future work. 
These are personal assessments on the most promising opportunities that may be lurking ahead of us.
In order to perform my task I start with a ``drone view'' of the current status of neutrino oscillations.
Here the major highlight has been the discovery of the phenomenon, made in solar and atmospheric studies, followed by reactor and accelerator-based experiments that have not only provided independent confirmation, but also improved parameter determination.
The current experimental data mainly converge to a consistent global picture in which the oscillation parameters are determined as shown in Fig.~\ref{fig:osc}. 

The top-three panels show the determination of the three mixing angles, where one sees that two of them are large, at odds with the corresponding mixing angles 
observed in the quark sector.
In fact, it is the smaller of the lepton mixing angles the one that lies intriguigly close in magnitude to the largest of the quark mixing angles, namely the Cabibbo angle.
Perhaps nature tries to tell us a secret right there.
One also sees that, although the overall picture of such ``three-neutrino paradigm'' is rather appealing, there are still some loose ends. 
The normal-ordered neutrino spectrum is preferred by slightly more than 3$\sigma$, while the octant of the atmospheric angle and the value of the leptonic CP phase are still poorly determined.

The ordering of the neutrino mass spectrum, as well as CP determination and octant resolution remain open challenges that 
form the target of the next generation of oscillation searches.
CP determination will be the task of the Deep Underground Neutrino Experiment (DUNE). 
The experiment will consist of two detector systems placed along Fermilab's Long Baseline Neutrino Facility (LBNF) beam~\cite{Acciarri:2015uup}. 
One detector system will be near the beam source, while a much larger one will be placed underground 1300 km away at the Sanford Underground Research Laboratory in South Dakota, in the same mine where Ray Davis has pioneered our field in the sixties.

\begin{figure}[h]
    \centering
\includegraphics[height=4.5cm,width=0.3\textwidth]{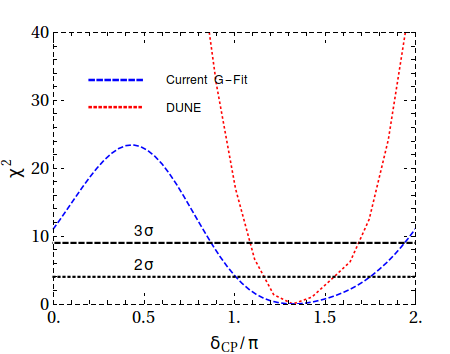}
\includegraphics[height=4.5cm,width=0.6\textwidth]{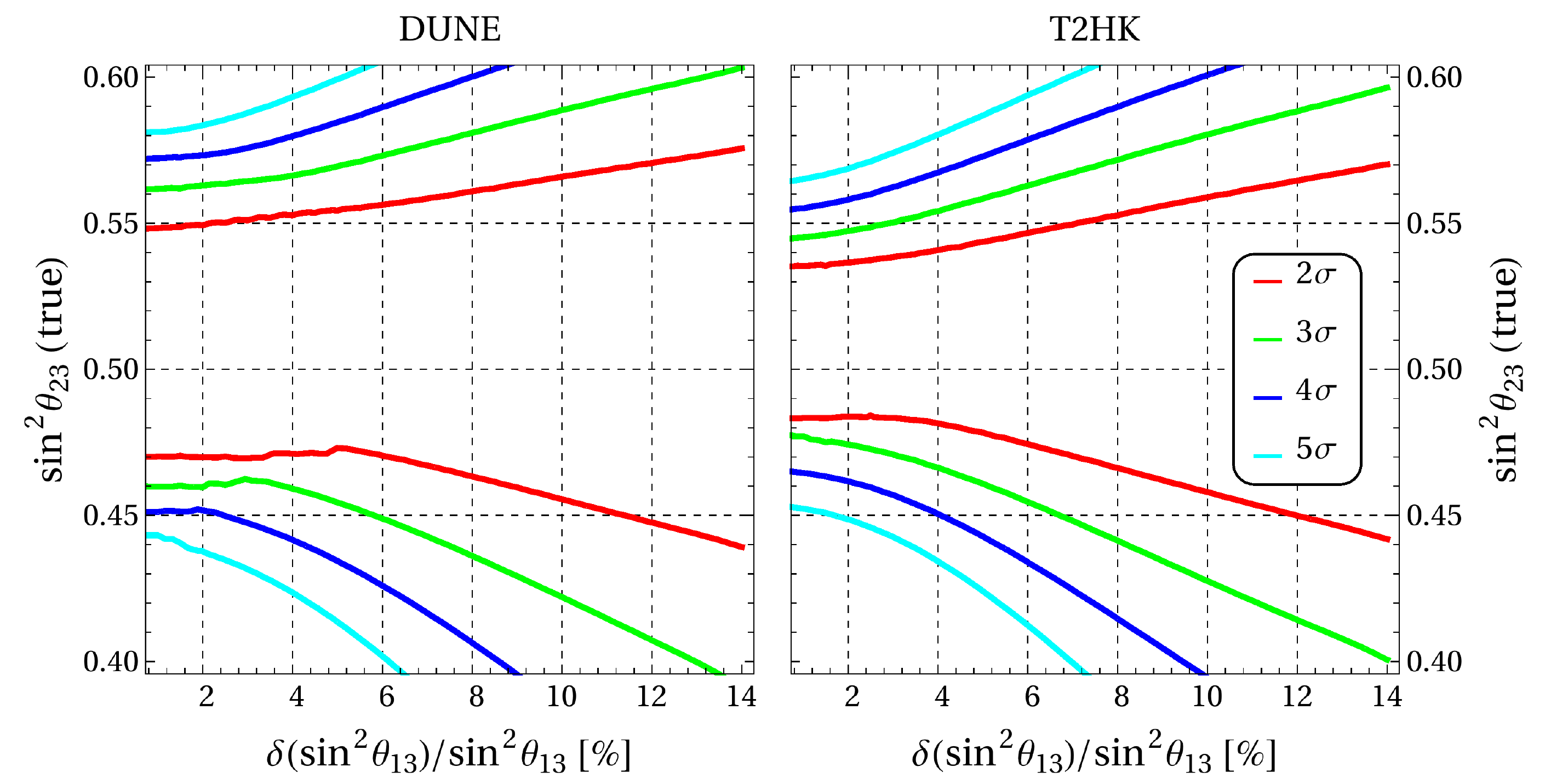}
\caption{Propects for CP measurement and octant determination. From~\cite{Nath:2018fvw,Chatterjee:2017irl}.}
    \label{fig:loose}
  \end{figure}

The situation concerning the CP phase is illustrated in Fig.~\ref{fig:loose}.
  The current status of CP is shown in blue in the left panel, as determined from the Valencia global fit (there is very good agreement amongst the three 
global fitting groups), while the future prospects are shown in red.
  The latter assume that the best fit value of the phase remains as given now, and that DUNE performs according to design.
  In this case the experiment will suffice to establish CP violation within the three-neutrino paradigm.
  In contrast, the other panels show that the octant resolution will not be fully achieved even by the ambitious T2HK proposal in Japan.
  See Ref.~\cite{Nath:2018fvw,Chatterjee:2017irl,Srivastava:2018ser} for details and related discussions.

\begin{figure}[h]
    \centering
\includegraphics[height=4cm,width=0.3\textwidth]{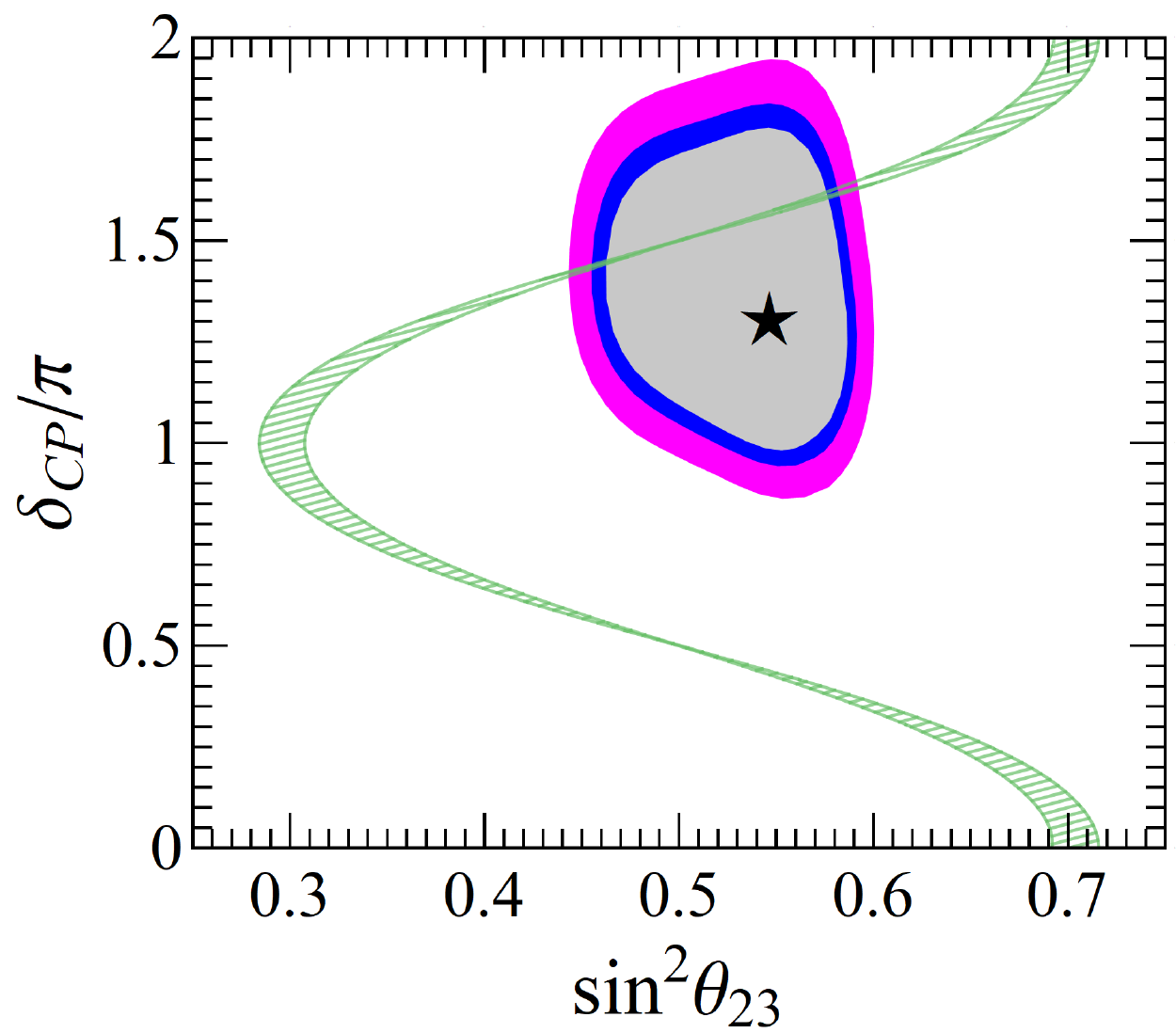}  
\caption{ Predicting the CP phase in a realistic generalized TBM scenario, from~\cite{Chen:2018eou}.}
    \label{fig:tbm}
  \end{figure}

Nevertheless, the two decades that followed the oscillation discovery have brought a tremendous progress in the determination of the pattern of leptonic mixing!
Its most salient features are nicely captured by the tri-bimaximal (TBM) mixing ansatz~\cite{Harrison:2002er}.
However, the TBM predictions are now at odds with observations from reactor experiments~\cite{An:2012eh}, and with CP violation hints from long-baseline oscillation experiments, such as T2K~\cite{Abe:2013hdq}.
Fortunately, there are systematic ways of generalizing patterns with manifest $\mu\tau$ reflection symmetry, such as TBM.
These ``revamped'' patterns are obtained by exploiting partially conserved generalized CP symmetries~\cite{Chen:2015siy,Chen:2016ica,Chen:2018lsv}, and their predictions may be tested with precision studies at DUNE~\cite{Nath:2018fvw}.
Indeed, as an example, the CP phase can be predicted to lie along the green band in Fig.~\ref{fig:tbm}~\cite{Chen:2018eou}. 
One sees how the overlap of the predicted band with the 3-neutrino global oscillation region selects a narrow allowed range for $\delta_{CP}$ in the upper branch.
Note however that, at least in this particular case, the octant determination challenge remains pretty much unresolved. 

\begin{figure}[h]
    \centering
\includegraphics[height=4.5cm,width=0.35\textwidth]{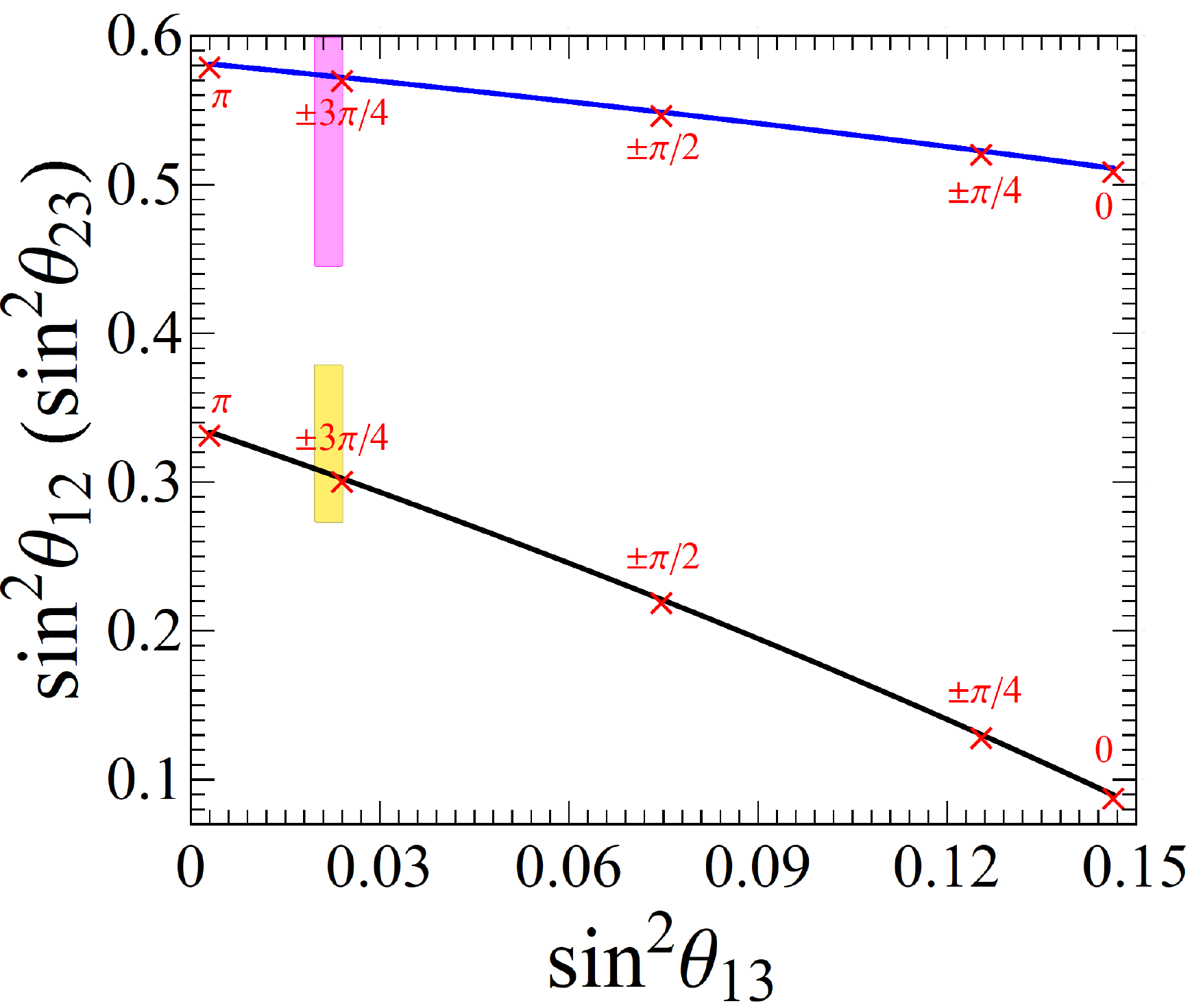}~~~~~
\includegraphics[height=4.5cm,width=0.35\textwidth]{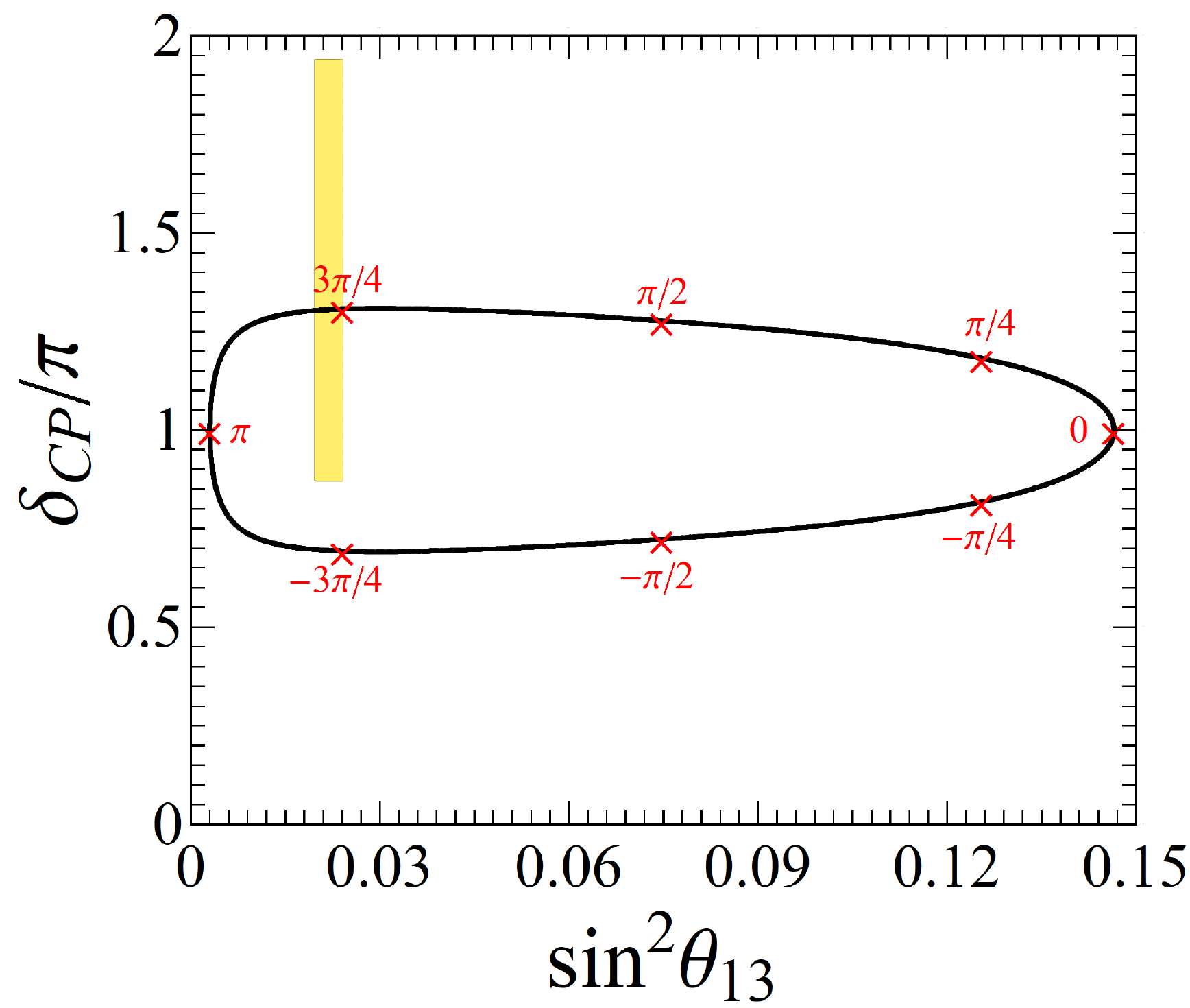}
\caption{Predictions for mixing angles and CP violation in the ``bi-large'' scheme of Ref.~\cite{Chen:2019egu}.}
    \label{fig:bl1}
  \end{figure}

There are other alternatives to the TBM pattern of neutrino mixing, for example bi-large mixing schemes~\cite{Boucenna:2012xb,Ding:2012wh}.
These exploit the observation made above that the largest quark-mixing is similar in magnitude to the smallest of the lepton mixing angles,
suggesting that the Cabibbo angle may be taken as the universal seed for flavor mixing~\cite{Roy:2014nua}.
Remarkably, in such ``bi-large'' lepton mixing schemes the good measurement of $\theta_{13}$ at reactor experiments allows one to predict \textit{both} solar and atmospheric mixing 
angles~\cite{Chen:2019egu} pretty well, as seen in Fig.~\ref{fig:bl1}, left panel. This also leads to sharp predictions for the CP phase, up to the degeneracy seen in the right panel, where two symmetric branches can be appreciated.

\begin{figure}[h]
    \centering
\includegraphics[height=5cm,width=0.5\textwidth]{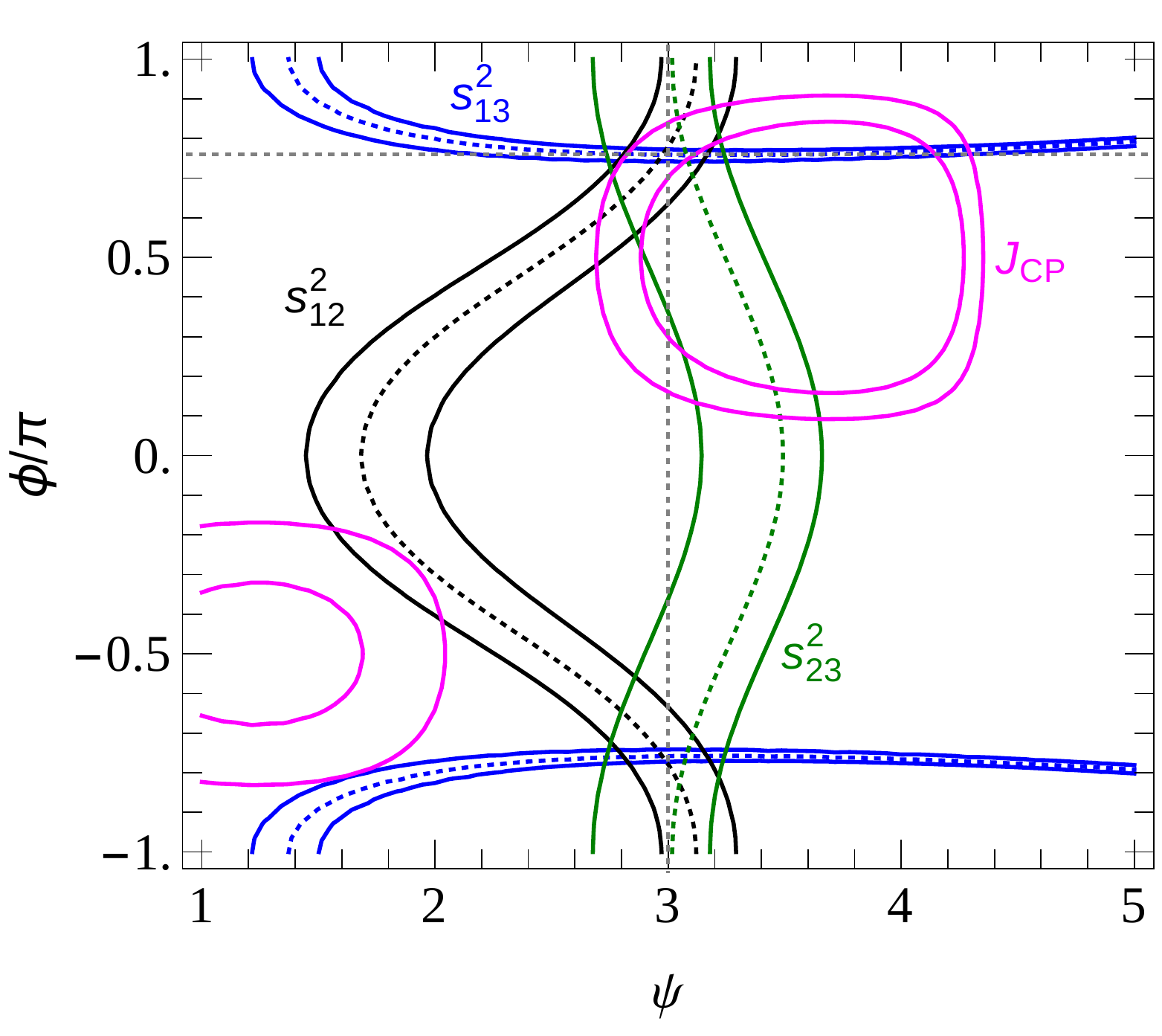}
\caption{Predictions for mixing angles and CP violation in a generalized ``bi-large'' scheme, from~\cite{Ding:2019vvi}.}
    \label{fig:bl2}
  \end{figure}

There are also ``softer'' bi-large-type lepton mixing patterns in which the oscillation parameters are expressed in terms of two independent parameters, $\phi$ and $\psi$, as seen in Fig.~\ref{fig:bl2}.
One sees how in this case the individual determinations of the three angles and the phase nicely converge into a unique region in the upper plane, indicating the utility of the bi-large description.

\begin{figure}[h]
    \centering
\includegraphics[height=4cm,width=0.4\textwidth]{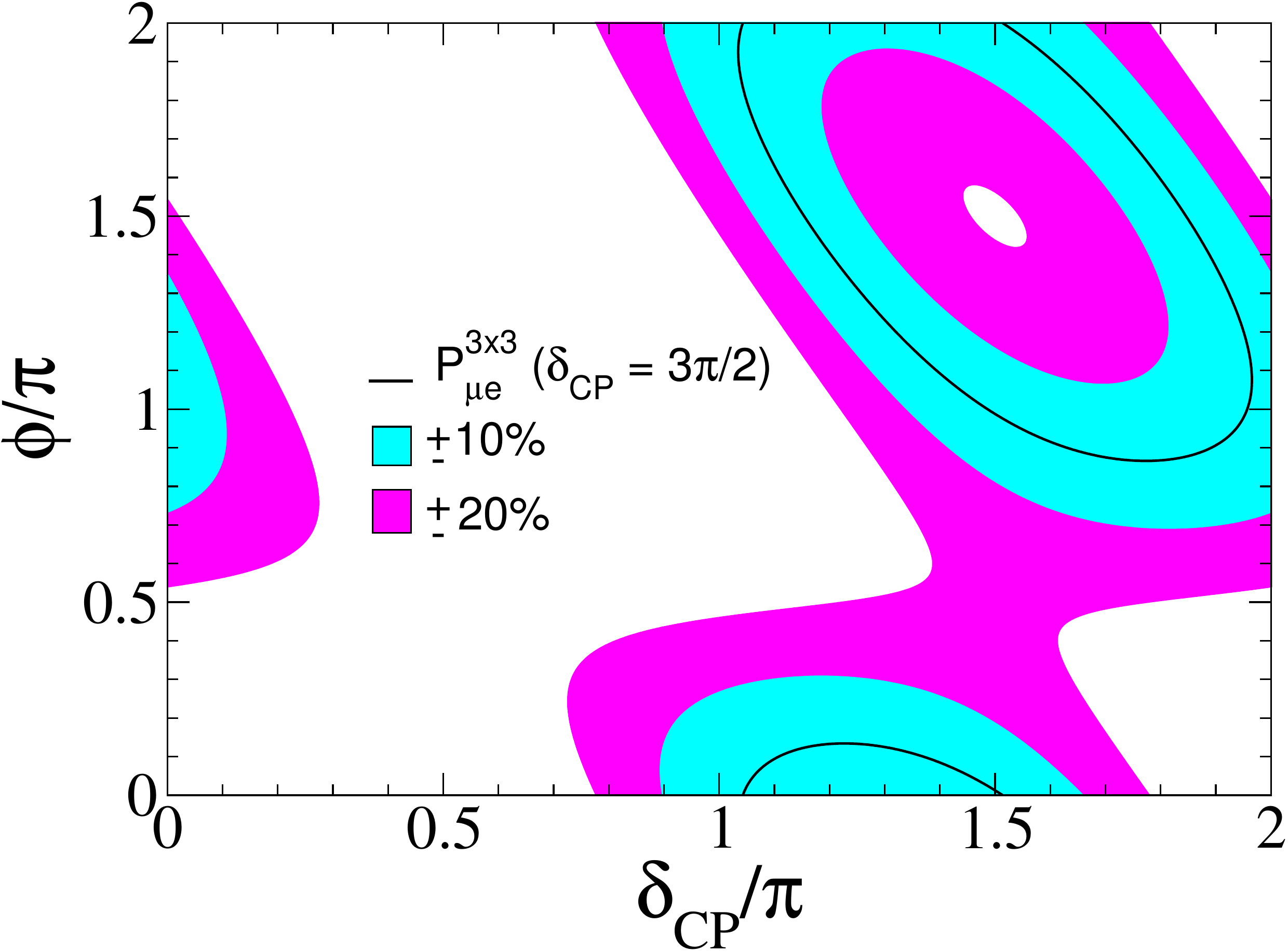}
\caption{New ambiguity in probing CP violation in neutrino oscillations, from Ref.~\cite{Miranda:2016wdr}. }
    \label{fig:confusion}
  \end{figure}

This brings us now to discuss the issue of robustness of the oscillation interpretation of neutrino data.
If neutrinos acquire mass from the exchange of heavy neutral isosinglet lepton mediators, as in the simplest seesaw paradigm, then these (mass-eigenstates) will couple (sub-dominantly) by mixing in the leptonic charged current~\cite{Schechter:1980gr}.
Since these states are too heavy to take part in oscillations, the mixing matrix describing the propagation of the three light neutrinos will not be strictly 
unitary~\cite{Valle:1987gv,Nunokawa:1996tg,Antusch:2006vwa,Miranda:2016ptb,Escrihuela:2015wra}.
The effective $3\times3$ mixing matrix will be written as,
\begin{equation}
K_L = N^{NP}~~U_{{\rm }},\label{eq:s-5}
\end{equation}
where the unitary matrix $U$ is multiplied by a pre-factor triangular matrix given as
\begin{equation}
 N^{NP} = \left(\begin{array}{ccc}
\alpha_{11} & 0 & 0\\
\alpha_{21} & \alpha_{22} & 0\\
\alpha_{31} & \alpha_{32} & \alpha_{33}
\end{array}\right).\label{eq:s-6}
\end{equation}
Its diagonal elements are real and close to one, while the off-diagonals are small but complex. 
It has been shown that the phase in $\alpha_{21}$ plays a crucial role in the CP measurement at DUNE. 
Indeed, there is an intrinsic ambiguity in probing CP violation in neutrino oscillations, arising from the confusion between the 
phase in $\alpha_{21}$ with that which characterizes the three-neutrino paradigm~\cite{Miranda:2016wdr}.
For example, for L/E = 500~km/GeV the vacuum appearance probability exhibits a degeneracy for different phase combinations.
This shows up in the  $P_{\mu e}$ isocontours as a function of the two CP phases, as indicated in Fig.~\ref{fig:confusion}.
The solid line corresponds to the standard value $P_{\mu e}$ with $\delta_{CP}=3\pi/2$, while colored regions denote the corresponding 10\%
and 20\% deviations.

\begin{figure}[h]
    \centering
\includegraphics[height=4cm,width=0.4\textwidth]{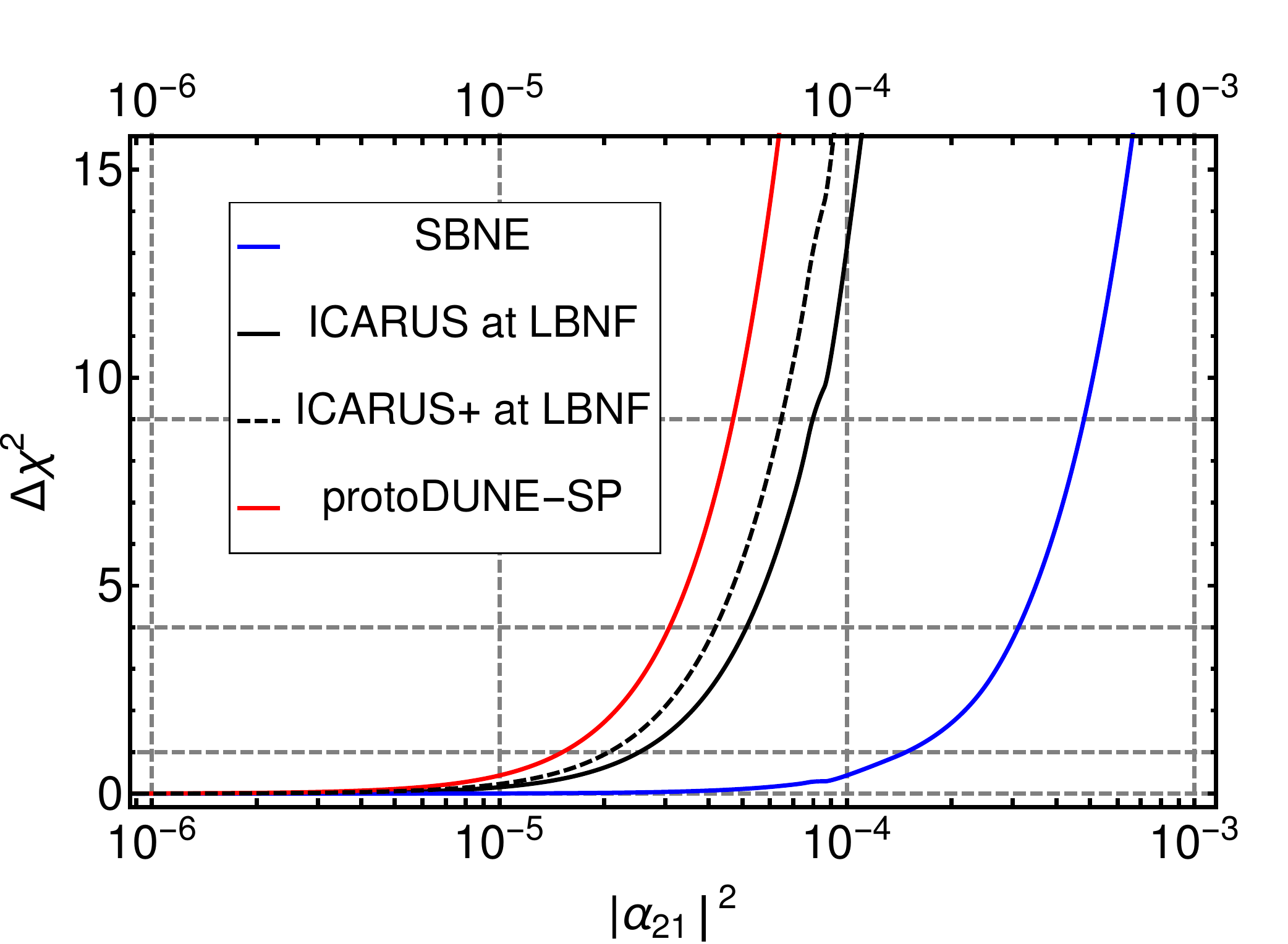}
\caption{Improved leptonic unitarity tests at various Fermilab near-distance setups, from Ref.~\cite{Miranda:2018yym}. }
    \label{fig:unitarity-violation-fermilab}
  \end{figure}

There have been suggestions to mitigate this problem~\cite{Ge:2016xya}, whose main ingredient is near detection of neutrinos,
capable of improving the limits on the magnitude of $\alpha_{21}$. 
The current bounds on the latter lie at the per-cent level and have been carefully compiled in Ref.~\cite{Escrihuela:2016ube}, 
stressing the robustness of neutrino-only limits.
The short-baseline neutrino program at Fermilab is ideally suited to probe the unitarity of the lepton mixing matrix, with meaningful sensitivities, 
potentially better than existing bounds arising from current neutrino experiments~\cite{Miranda:2018yym}, as illustrated in Fig.~\ref{fig:unitarity-violation-fermilab}.

It would be desirable to perform more dedicated experimental studies of the potential of various near-distance setups, taking into account their practical feasibility.

Needless to say that unitarity violation is only the simplest example of non-standard neutrino interactions (NSI). 
These are a generic feature of low-scale models of neutrino mass generation~\cite{Boucenna:2014zba}, and their study constitutes an important topic in 
neutrino physics~\cite{Maltoni:2004ei} as it could shed light on the scale of neutrino mass generation and probe the robustness of the neutrino oscillation 
interpretation~\cite{Miranda:2004nb}. NSI could also bring degeneracies in the determination of the CP phase, similar to what we just discussed~\cite{Flores:2018kwk}.
 As a result, NSI studies constitute a necessary topic in the agenda of upcoming oscillation experiments~\cite{Dev:2019anc}.

\section{Absolute neutrino mass} 

Although the oscillation programme has driven much of the ``revolution'' our field has experienced over the last few decades, 
by themselves, oscillation studies are insensitive to the absolute scale of neutrino mass.
Single and double beta decay processes (as well as cosmology) can probe the absolute neutrino mass in complementary ways.

Recently the Katrin experiment has derived an upper limit of 1.1 eV (at 90\% C.L.) on the absolute mass scale of 
neutrinos~\cite{Aker:2019uuj} from the study of Tritium endpoint spectrum. 
This bound applies irrespectively of whether neutrinos are Dirac or Majorana particles.
In the latter case one expects also a neutrinoless variety of double beta decay -- dubbed \znbb for short -- in which no neutrinos are emitted as real particles.
Rather, the process involves the virtual \lnv propagation of neutrinos.

\begin{figure}[h]
    \centering
\includegraphics[height=4.5cm,width=0.45\textwidth]{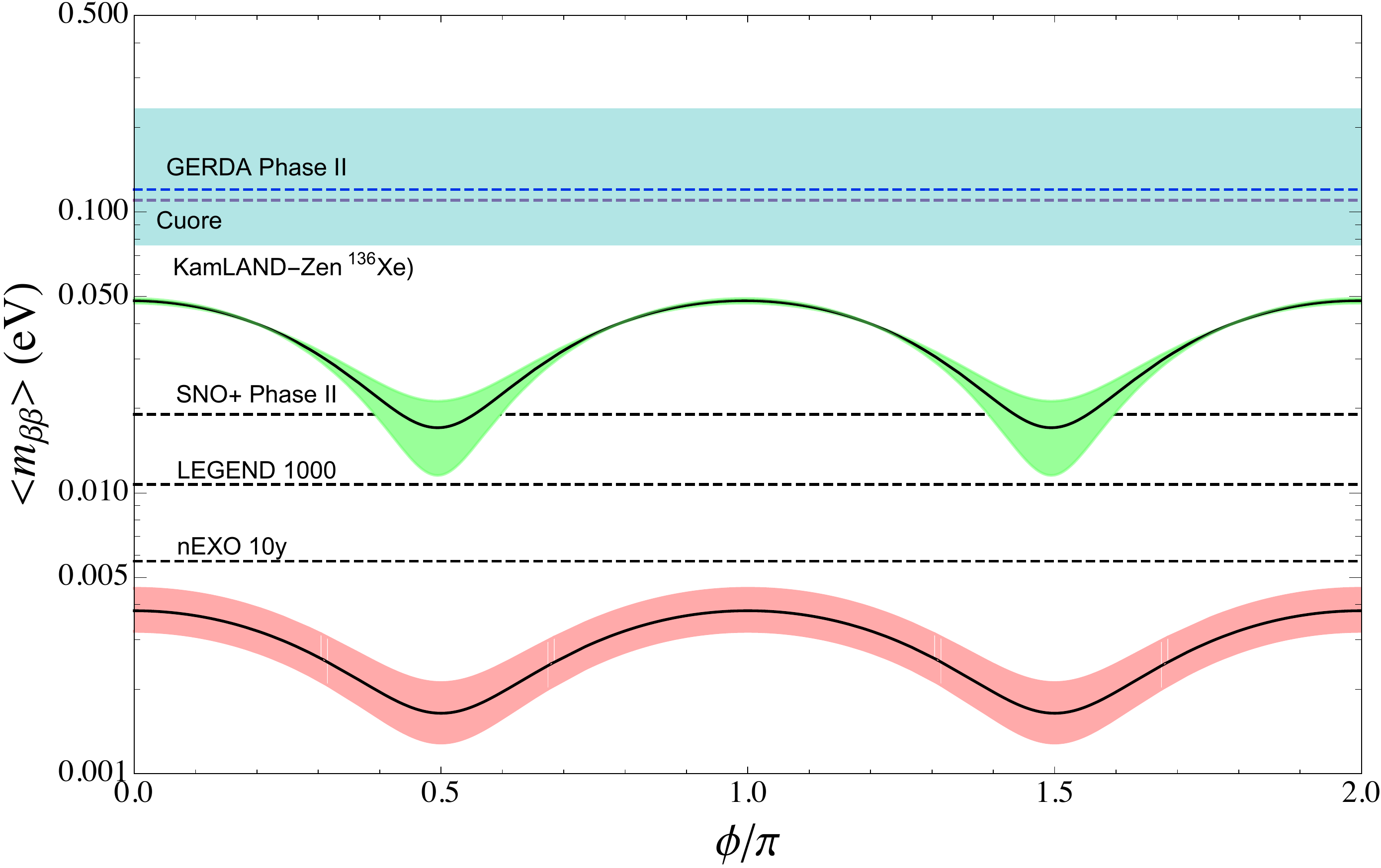}~~~~~
\includegraphics[height=4.5cm,width=0.45\textwidth]{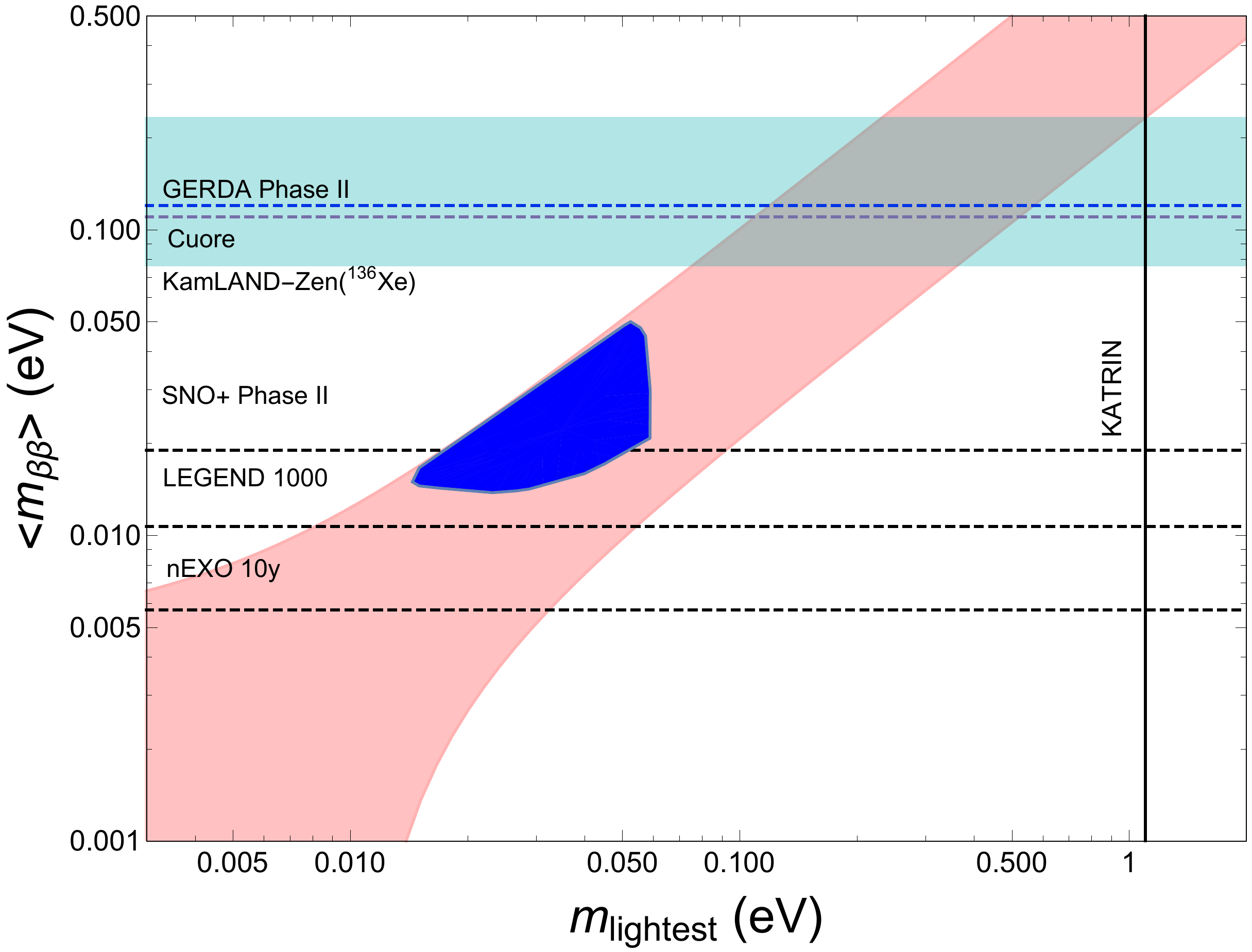} 
\caption{Lower bound on the \znbb decay amplitude in a two-massive-Majorana-neutrino model, e.g.~\cite{Reig:2018ztc},
(left), and in the flavor-symmetric three-massive-Majorana-neutrino model of~\cite{King:2013hj} (right). }
    \label{fig:dbd}
  \end{figure}
One can determine the expected ranges for the decay amplitude, taking into account the allowed neutrino 
oscillation parameters given above. 
One finds that for inverted-ordered neutrino masses, there is a lower bound for the \znbb amplitude, 
while for normal-ordered neutrinos there is none, indicating the possibility of having full destructive 
interference amongst the three light neutrinos.
This is of course discouraging for experiment. 

However, if one of the three neutrinos is massless or nearly so, there is no cancellation, 
even if neutrinos are normal-ordered. 
Note that the resulting lower bound correlates with the only free parameter, the relative Majorana phase 
between the two neutrinos, see left panel in Fig.~\ref{fig:dbd}. As a result, the allowed \znbb ranges
are also much narrower than in the generic three-neutrino case.
This situation emerges in a number of theories, such as the original missing partner seesaw 
mechanism~\cite{Schechter:1980gr} and many other schemes, for example the one considered in~\cite{Reig:2018ztc}.

Even if all three neutrinos are massive, it could happen that cancellation is prevented by the
flavor structure of the leptonic weak interaction vertex, as predicted by a number of flavor
symmetries~\cite{Dorame:2011eb,Dorame:2012zv}.
Either way, there is hope that upcoming experiments may shed light on the relevant Majorana phase,
as indicated by the estimated experimental
sensitivities~\cite{KamLAND-Zen:2016pfg,Alduino:2017ehq,Albert:2017owj,Agostini:2018tnm,Andringa:2015tza,Abgrall:2017syy,Albert:2017hjq}.
For an alternative recent compilation of \znbb sensitivities see~\cite{Agostini:2019hzm}.

To sum up the \znbb discussion, we now recall the old black-box theorem~\cite{Schechter:1981bd}, 
revisited in Ref.~\cite{Duerr:2011zd}.
That argument captures the significance of the \znbb decay by stating that, if this process is ever 
discovered, it would imply the Majorana nature of (at least one of) the neutrinos.
The conclusion holds irrespective of whether \znbb arises from neutrino exchange or from a short-range mechanism 
that might be probed at colliders, such as the LHC.

\section{Origin of neutrino mass}

Forty years ago Weinberg noted that, even though the \sm lacks neutrino masses, one can induce them 
through a unique dimension five operator~\cite{Weinberg:1979sa} associated to new physics with lepton 
number non-conservation. 
The most popular ``UV-completion'' of the dimension five operator is provided by the seesaw mechanism.
Here neutrino masses are induced by the exchange of heavy fermions (type I) or scalars (type II), 
a terminology which is opposite to the one adopted in~\cite{Schechter:1980gr}.
The new seesaw states were originally thought to lie at a high mass scale, associated to 
unification, whose most characteristic predictions (proton decay) have so far not been vindicated.

Here we focus instead on ``low-scale'' realizations of the seesaw mechanism~\cite{Mohapatra:1986bd,Akhmedov:1995vm,Akhmedov:1995ip,Malinsky:2005bi}.
Notice that, at the standard \SM level, ``right-handed'' neutrinos are gauge-singlets, 
so their number is arbitrary, it need not match the number of ``left-handed'' ones~\cite{Schechter:1980gr}.

Hence one can add less ``right-'' than ``left-handed'' neutrinos. As an example, one may consider a (3,2) seesaw scheme, 
with three ``left-'' and only two ``right'' neutrinos. In this case one of the ``left-handed'' neutrinos remains ``unpaired'' 
and hence massless~\cite{Schechter:1980gr}. 
In such ``missing partner seesaw'' both solar and atmospheric scales arise from the seesaw mechanism, 
and the lightest neutrino remains massless, leading to the \znbb lower bound we just discussed.

Likewise, a (3,1) seesaw scheme can be envisaged. 
In this case only one mass scale is generated by the tree-level seesaw mechanism, 
while the other, the solar scale, may arise from some loop mechanism mediated, for example, 
by supersymmetry~\cite{Hirsch:2000ef,Diaz:2003as}, 
or by a ``dark sector'', as in the simplest scotogenic seesaw mechanism~\cite{Rojas:2018wym}.

Alternatively, one may add more ``right-'' than ``left-handed'' neutrinos. As a very interesting example, 
one can add two isosinglets per family of leptons, sequentially. 
By imposing lepton number conservation on such (3,6) scheme (again, in the same notation of~\cite{Schechter:1980gr})
one gets a ``template'' scheme with massless neutrinos, exactly as in the Standard Model.
In other words, such simple setup leads to massless neutrinos within perturbation theory, as long as lepton number 
symmetry is exact.
In contrast to the \sm case, however, lepton flavor is violated, and similarly, CP symmetry. 
This has two important implications. First, it elucidates the meaning of flavor and CP violation
in the leptonic weak interaction, implying that such processes need not be suppressed by the 
smallness of neutrino masses, and can therefore be large~\cite{Bernabeu:1987gr,Branco:1989bn,Rius:1989gk,Deppisch:2004fa,Deppisch:2005zm}.
Second, this reference model serves also as template for building genuine low-scale seesaw schemes 
where neutrino masses are protected by lepton number symmetry~\cite{Mohapatra:1986bd,Akhmedov:1995vm,Akhmedov:1995ip,Malinsky:2005bi}.

In short, neutrino masses are naturally small as a result of symmetry protection both in the high-scale as well as low-scale seesaw approaches.
One expects a rich phenomenology in low-scale seesaw, in contrast to the high-scale seesaw mechanism. 

We now turn to the possibility of Dirac neutrinos. Whether or not neutrinos are Dirac-type is an experimental open question.
If neutrinos happen to be Dirac particles, symmetry is required not only to
account for the small neutrino masses, but also to ensure \textit{Diracness}.
In fact, there might be a deep reason for neutrinos to be Dirac-type. For example, this could be associated
to the stability of dark matter~\cite{Chulia:2016ngi,CentellesChulia:2017koy} or to the existence of a Peccei-Quinn symmetry~\cite{Peinado:2019mrn}.
In the last few years there have been detailed classifications of Dirac seesaw 
mechanisms~\cite{CentellesChulia:2018gwr,CentellesChulia:2018bkz} as well as full-fledged UV-complete model 
constructions~\cite{Chulia:2016ngi,CentellesChulia:2017koy,Reig:2016ewy,Valle:2016kyz,Bonilla:2016zef,Chulia:2016giq}.

An important point to note is that the seesaw opens the way for a \textit{dynamical}
understanding of small neutrino masses where, in addition to the standard vacuum 
expectation value (vev) $v_2$ responsible for electroweak breaking, there are new vevs coupled
to neutrinos.
These include an isotriplet $v_3$ coupled to ``left-nanded'' neutrinos, and an isosinglet 
$v_1$ coupled to ``right'' neutrinos. These may trigger the spontaneous violation of lepton
number, accompanied by a Goldstone boson, dubbed Majoron~\cite{Chikashige:1980ui}.
The dynamical \SM seesaw mechanism with hierarchical vevs  $v_{1}\gg v_{2}\gg v_{3}$ was proposed in~\cite{Schechter:1981cv}.
It has been noted that the extra vevs can substantially affect and improve
vacuum stability and perturbative unitarity within low-scale seesaw schemes~\cite{Bonilla:2015eha,Bonilla:2015kna}.
 
\section{Probing the mediators of neutrino mass generation}

For definiteness, here we take the simplest low-scale type-I seesaw mechanism of neutrino 
mass generation~\cite{Mohapatra:1986bd,Akhmedov:1995vm,Akhmedov:1995ip,Malinsky:2005bi}.
As we saw, if the seesaw mediators lie in the TeV scale, they can lead to a variety of observable effects.
For example, their existence will affect the description of neutrino oscillations, by having 
an effectively non-unitary mixing matrix describing the propagation of the three light neutrinos ~\cite{Valle:1987gv,Nunokawa:1996tg,Antusch:2006vwa,Miranda:2016ptb,Escrihuela:2015wra}.
As we have already discussed, this brings in extra CP violation that can fake the one expected within the simplest three-neutrino 
paradigm~\cite{Miranda:2016wdr}.
As a result, unitarity violation degrades the CP violation sensitivity expected at DUNE~\cite{Escrihuela:2016ube}.


Low-scale seesaw models also open the possibility of direct production of the mediators of neutrino mass generation in collider experiments.
Since the neutrino mass mediators in type-I seesaw are gauge singlets, they couple to the weak currents 
only through mixing~\cite{Schechter:1980gr}.
Yet, already in the pre-LEP days, it was suggested that the isosinglet heavy leptons $N$ present 
in type-I seesaw would be singly-produced in Z boson decays, as $Z \to N + \nu$, where $\nu$ 
is a light neutrino~\cite{Dittmar:1989yg}.
Given the large number of Z's at the peak, the detection of the associated signatures would be 
feasible all the way up to masses close to the Z mass~\cite{Dittmar:1989yg}.
The Delphi and L3 collaborations have later performed this search at the Z-peak and above.
In the latter case one covers a wider kinematical range, but the sensitivity worsens due to the lower rates.

Proton-proton collisions at the LHC would singly-produce the $N$ by a Drell-Yan-type mechanism,
and searches have been performed by ATLAS and CMS, further extending the kinematical reach. 
Again, as a result of the low single-production rates, the sensitivity is not as high.

\begin{figure}[h]
    \centering
\includegraphics[height=5.5cm,width=0.45\textwidth]{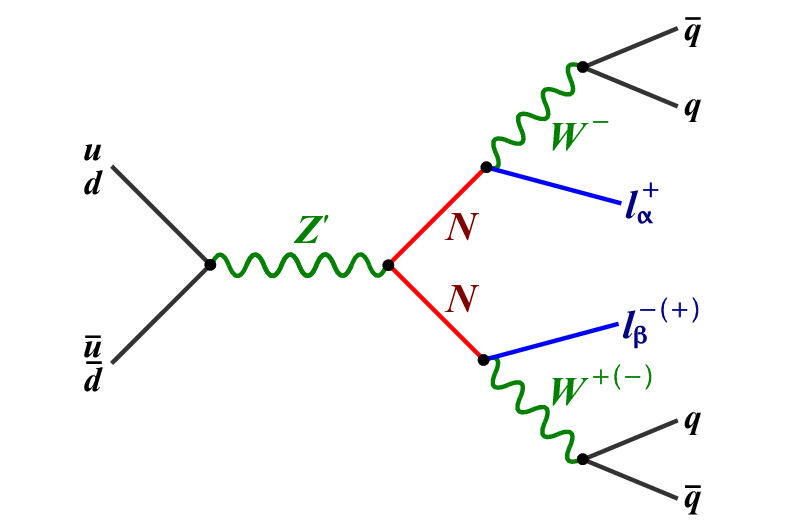}~~~~~
\includegraphics[height=5.5cm,width=0.45\textwidth]{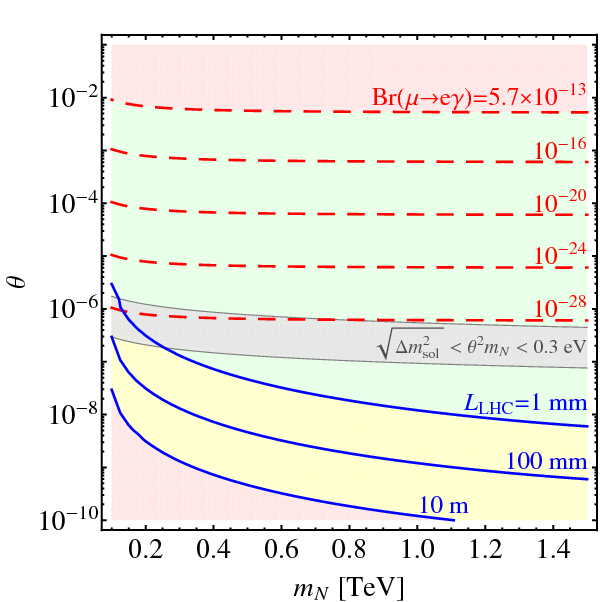}
\caption{Average decay length of a heavy neutrino $N$ produced via a $Z^\prime$
decaying as $Z^\prime \to NN$, versus the mass $m_N$ and the light-heavy mixing angle $\theta$ (solid blue contours), from~\cite{Deppisch:2013cya}, see text.}
    \label{fig:clfv}
  \end{figure}

A way to avoid the mixing suppression is to imagine the existence of a production portal provided by new 
vector bosons associated to an extended gauge symmetry~\cite{Das:2012ii,Deppisch:2013cya,Deppisch:2015qwa}.
In this case, the exchange of a new $Z^\prime$ can pair-produce the $N$ \textit{a la Drell-Yan}, 
leading to improved sensitivities~\cite{Das:2012ii,Deppisch:2013cya,Deppisch:2015qwa}.
Most interestingly, each $N$ can decay to any flavor of charged leptons.
This would lead to a unique possibility of \clfv at high energies, proposed in~\cite{Bernabeu:1987gr}.
Moreover, since the $N$ decays through mixing, the associated events may show a displaced vertex, as seen in Fig.~\ref{fig:clfv}, right panel.

The plot shows the average decay length of a heavy neutrino $N$ produced via a 3 TeV mass $Z^\prime$
decaying as $Z^\prime \to NN$, versus $m_N$ and the light-heavy mixing angle $\theta$ (solid blue contours). 
The dashed-red contours denote constant values for $BR(\mu\to e+\gamma)$, whereas the grey shaded band 
corresponds to parameter values which produce light neutrino mass scales around 0.3~eV within the simplest
type-I seesaw mechanism.
One sees that, for some parameter values, the sensitivity attainable at high energies
will exceed that of low-energy \lfv processes such as $\mu\to e+\gamma$.
This nicely illustrates the complementarity between high energy colliders and the high intensity 
approach pursued at muon facilities.
Moreover, it provides a real possibility for an important (double) discovery at the LHC, i.e. that of the mediator
of neutrino mass generation, as well as the discovery of the \clfv phenomenon at high energies!

\section{Conclusion}

All in all, the legacy of the oscillation programme over the last two decades has been a tremendous progress in our field, 
bringing neutrinos to the center of the particle physics stage.
Indeed, addressing the dynamical origin of small neutrino masses touches the heart of the electroweak theory, such as the consistency of symmetry breaking.
Besides neutrino mass dynamics, there are other issues in particle physics and cosmology for which neutrinos may provide key input.
For example, they could shed light into the flavor problem~\cite{Babu:2002dz,Morisi:2013qna,Chatterjee:2017ilf,Chen:2015jta}, 
give us a glimpse for the existence of extra dimensions~\cite{Chen:2015jta},
or suggest new pathways to unification~\cite{Boucenna:2014dia,Deppisch:2016jzl}.
Perhaps they could also make a step forward towards the final dream of unifying the existing forces and the oberved particle families together~\cite{Reig:2017nrz}.
Last but not least, dark matter and neutrinos could be intimately related, so neutrinos may, in some sense, also hold the key
to the solution of the cosmological dark matter problem.
For example, dark matter could be the mediator of neutrino mass generation~\cite{Ma:2006km,Avila:2019hhv}.
Unfortunately, I will not have the space to go into these in detail, see~\cite{Valle:2017fwa} for some comments.

From the experimental viewpoint in the coming decade we expect a vibrant period for oscillations studies, within and beyond the minimum paradigm.
Studies should also be performed using neutral current phenomena.
Likewise \znbb searches may lead to a breakthrough in the next decade or so.
High energy studies at the LHC may prove better for \clfv searches than the conventional high intensity muon facilities, 
and might have a chance to actually discover the mediators ultimately responsible for neutrino mass generation! \\

I take the opportunity to thank all of my collaborators for the good moments we have shared doing neutrino physics. 
In particular I thank O. Miranda for going through the text. 


\end{document}